\documentstyle[pre,multicol,aps]{revtex}

\input{epsf.tex}
\newcommand{\nem}{\em}
\newcommand{\pt}[1]{{\em ``#1,'' }}   

\begin{document}
\title{Nonlinear guided waves and spatial solitons in a periodic layered medium}

\author{ Andrey A. Sukhorukov and Yuri S. Kivshar} 

\address{Nonlinear Physics Group, Research School of Physical Sciences and Engineering, Australian National University,\\ 
Canberra ACT 0200, Australia}
\maketitle

\begin{abstract}
We overview the properties of nonlinear guided waves and (bright and dark) spatial optical solitons in a periodic medium created by a sequence of linear and nonlinear layers. First, we consider a single layer with a cubic nonlinear response (a nonlinear waveguide) embedded into a periodic layered linear medium, and describe nonlinear localized modes (guided waves and Bragg-like localized gap modes) and their stability.  Then, we study modulational instability as well as the existence and stability of discrete spatial solitons in a periodic array of identical nonlinear layers, a one-dimensional {\em nonlinear photonic crystal}. Both similarities and differences with the models described by the discrete nonlinear Schr\"odinger equation (derived in the tight-binding approximation) and coupled-mode theory (valid for the shallow periodic modulations) are emphasized.
\end{abstract}

\begin{multicols}{2}
\narrowtext
\section{Introduction}

Spatially localized waves (or intrinsic localized modes) in nonlinear lattices have been an active research topic during last years. In application to the problems of nonlinear optics, such modes are known as {\em discrete spatial solitons}, and they have been described theoretically~\cite{theory,theory1} and recently observed experimentally~\cite{exp} in periodic arrays of nonlinear single-mode optical waveguides. A standard approach in the study of the discrete spatial solitons in optical super-lattices is to employ the properties of an effective discrete nonlinear Schr\"odinger (NLS) equation, which can be derived under some assumptions, similar to the tight-binding approximation in the solid-state physics~\cite{theory}. On the other hand, nonlinear localized waves in a system with a weakly modulated optical refractive index are known as {\em gap solitons}~\cite{gap_review}. The similar problems and methods of their solutions appear in other fields, such as the nonlinear dynamics of the Bose-Einstein condensates in optical lattices~\cite{smerzi}.

However, real experiments in guided-wave optics are conducted in the structures of more complicated geometries and, therefore, the applicability of the tight-binding approximation and the corresponding discrete equations become questionable. Moreover, one of the main features of the wave propagation in periodic structures (which follows from the Floquet-Bloch theory) is the existence of forbidden transmission band gaps and, therefore, the nonlinearly-induced wave localization can be also possible in the form of the so-called {\em gap solitons} located in each of these gaps. However, the effective discrete equations derived in the tight-binding approximation describe only one transmission band surrounded by two semi-infinite band gaps and, therefore, the real fine structure of the band-gap spectrum associated with the wave transmission in a periodic medium is lost. On the other hand, the coupled-mode theory of the gap solitons~\cite{gap_review} describes only the modes localized in an isolated narrow gap, and it does not allow to consider simultaneously the gap modes and conventional guided waves localized due to the total internal reflection. However, the complete band structure of the transmission spectrum and simultaneous existence of localized modes of different types is very important in the analysis of the stability of nonlinear localized modes. Such an analysis is especially important for the theory of nonlinear localized modes and nonlinear waveguides in realistic models of nonlinear photonic crystals (see, e.g., the recent paper~\cite{PBG}, and references therein).

In this paper, we consider a simple model of a nonlinear periodic layered medium where the optical super-lattice is formed by a periodic sequence of two linear layers of different dielectric susceptibilities, and nonlinear waveguides are described by thin-film layers embedded into it~\cite{our_pre} (see also Ref.~\cite{gera}). In such a case, the effects of the linear periodicity and band-gap spectrum structure are taken into account explicitly, and nonlinearity enters the corresponding matching conditions only, allowing a direct analytical study.

Here, we consider two problems of this kind. In the first case (Sec.~\ref{sect:layer}), we study the nonlinear guided waves supported by a thin isolated nonlinear layer with the Kerr-type nonlinear response, which is embedded into a linear periodic structure composed of two layers with different linear dielectric constants. We consider the case when the layer describes an optical waveguide that supports guided waves in a homogeneous medium of the averaged susceptibility, and assume that the cubic nonlinear response may be either positive (self-focusing) or negative (self-defocusing). We describe two different types of nonlinear localized modes supported by this structure, including the analysis of the mode stability. In the second case (Sec.~\ref{sect:array}), we analyze the nonlinear localized modes in an infinite structure consisting of a periodic sequence of nonlinear layers, similar to the geometry of the experiments with discrete optical solitons~\cite{exp}. In this case, we study first modulational instability in both self-focusing and self-defocusing regimes, and then discuss the properties of different types of nonlinear localized modes such as bright, dark and ``twisted'' spatial solitons. We emphasize both similarities and differences with the models described by the discrete NLS equation and the continuum coupled-mode theory.

\section{Nonlinear guided modes\\ in a periodic medium}
         \label{sect:layer}
\subsection{Model} \label{sect:layer_model}

We consider the wave propagation along the $Z$-direction of a slab waveguide consisting of a sequence of linear and nonlinear layers.  The evolution for the complex electric field envelope $E(X,Z)$ in the waveguide is governed by the NLS  equation,
\begin{equation} \label{eq:nls_dim} 
       i \frac{\partial E}{\partial Z} 
       + D \frac{\partial^2 E}{\partial X^2} 
       + \varepsilon(X) E + g(X) |E|^2 E = 0,
\end{equation}
where $D$ is the diffraction coefficient ($D>0$). The phase velocity of the fundamental mode is defined by the function $\varepsilon(X)$, and $g(X)$ characterizes the Kerr-type nonlinear response. We assume that either the function $\varepsilon(X)$ or $g(X)$ (or both of them) is periodic in $X$, i.e. it describes the periodic layered structure similar to the so-called transverse Bragg waveguides created by the nonlinear thin-film multilayer structures~\cite{grebel,nabiev} or the impurity band in a deep photonic band gap~\cite{lan}.  

In order to reduce the number of physical parameters, we normalize Eqs.~(\ref{eq:nls_dim}) as follows: 
$E(X,Z) = \psi(x,z) E_0 e^{i \; \overline{\varepsilon} Z}$, where $\overline{\varepsilon}$ is the mean value of the function $\varepsilon(X)$, $x=X/d$ and $z=Z D / d^2$ are the dimensionless coordinates, $d$ and $E_0$ are the characteristic transverse scale and field amplitude, respectively. Then, the normalized nonlinear equation has the form
\begin{equation} \label{eq:nls}  
     i \frac{\partial \psi}{\partial z} 
     + \frac{\partial^2 \psi}{\partial x^2}  
     + {\cal F}(I; x) \psi = 0,
\end{equation}
where the real function ${\cal F}(I; x) = d^2 D^{-1} [ \varepsilon(X) - \overline{\varepsilon} + g(X) I |E_0|^2]$ describes both {\nem nonlinear} and {\nem periodic} properties of the layered medium, and $I \equiv |\psi|^2$ is the normalized local wave intensity. We note that the system~(\ref{eq:nls}) is Hamiltonian, and for the spatially localized solutions the power, 
\[
   P = \int_{-\infty}^{+\infty} |\psi(x,z)|^2 \; dx ,
\]    
is conserved.

At this point, it is important to mention that Eq.~(\ref{eq:nls}) describes the beam evolution in the framework of the so-called {\em parabolic approximation}, valid for the waves propagating mainly along the $z$ direction (see also Ref. \cite{our_old_pre}).  In other words, the characteristic length of the beam distortion due to both diffraction and refraction along the $z$ axis should be much  larger than the beam width in the transverse direction $x$. This leads to the condition of {\em a weakly modulated periodicity},  $|\varepsilon(X) - \overline{\varepsilon}| 
                                  \ll |\overline{\varepsilon}|$. 

We look for stationary localized solutions of the normalized equation (\ref{eq:nls}) in the form
\begin{equation} \label{eq:lmode}
 \psi(x,t) = u(x; \beta) e^{i \beta z}, 
\end{equation}
where $\beta$ is the propagation constant, and the real function $u(x; \beta)$ satisfies the stationary nonlinear equation:
\begin{equation} \label{eq:u0_inh}
  - \beta u + \frac{d^2 u}{d x^2} + {\cal F}(I; x) u = 0.
\end{equation}
We assume that the basic optical super-lattice is linear, and nonlinearity appears only through the properties of a thin-film waveguide (or an array of such waveguides) Then, if the corresponding width of the wave envelope is much larger than that of the waveguide, the thin-film waveguide can be modeled by a delta-function and, in the simplest case of a single nonlinear layer, we can write 
   ${\cal F}(I; x) = \nu(x) + \delta(x) G(I)$, 
where the function $G(I)$ characterizes the properties of the nonlinear thin-film waveguide, and $\nu(x) \equiv \nu(x+h)$ describes an effective potential of the superlattice with the spatial period $h$. 

For such a nonlinearity, localized waves can be constructed with the help of certain matching conditions, by using the solutions of Eq.~(\ref{eq:u0_inh}) with ${\cal F}(I; x) = \nu(x)$, presented in the form of the Bloch-type wave functions~\cite{yeh}. Additionally, this approximation can be easily extended to the case of a periodic array of thin-film nonlinear waveguides (see Sec.~\ref{sect:array_model} below). 

\subsection{Band-gap spectrum and localized waves}
            \label{sect:layer_bands}

If the effective periodic potential $\nu(x)$ is approximated by a piecewise-constant function (such a model is known as the Kronig-Penney model), the solution can be decomposed into a pair of counter-propagating waves with the amplitudes $a(x; \beta)$ and $b(x; \beta)$,
\begin{equation} \label{eq:bwaves}
  u_b(x; \beta) = a(x; \beta) e^{- {\mu}(x; \beta) x} 
         + b(x; \beta) e^{+ {\mu}(x; \beta) x}, 
\end{equation}
where $\mu(x; \beta) = \sqrt{\beta - \nu(x)}$ is the local wavenumber.
As follows from the Floquet-Bloch theory, for a Bloch-wave solution the reflection coefficient $r(x; \beta) = b(x; \beta) / a(x; \beta)$ is a periodic function, i.e. $r(x; \beta) = r(x + h; \beta)$, and it satisfies the eigenvalue problem:
\begin{equation} \label{eq:eigen}
   T(x; \beta) \left( \begin{array}{l} 1 \\ r(x; \beta) \end{array} \right) 
    =
      \tau(\beta) 
      \left( \begin{array}{l} 1 \\ r(x; \beta) \end{array} \right) ,
\end{equation}
where $T(x; \beta)$ is a {\nem transfer matrix} that describes a change of the wave amplitudes $\{a,b\}$ after one period $(x,x+h)$~\cite{transfer}. 
It was found that ${\rm det}\;T \equiv 1$ and, therefore, two linearly independent solutions of Eq.~(\ref{eq:eigen}) correspond to a pair of the eigenvalues $\tau$ and $\tau^{-1}$. Relation $\tau(\beta)$ determines a {\nem band-gap structure} of the superlattice spectrum: the waves are {\nem propagating}, if $|\tau| = 1$, and they are {\nem localized}, if $|\tau| \ne 1$. In the latter case, a nonlinear waveguide can support {\nem nonlinear localized waves} as bound states of Bloch-wave solutions with the asymptotics $|u( x \rightarrow \pm \infty )| \rightarrow 0$.
The wave amplitude at the thin-film nonlinear waveguide is determined from the continuity condition at $x=0$, i.e. $I_0 \equiv I(0^+) =  I(0^-)$ and $[d u / d x]_{x=0^-}^{0^+} + G(I_0) u_{x=0} =0$. Then, we use Eq.~(\ref{eq:bwaves}) to express the latter condition through the superlattice characteristics,
\begin{equation} \label{eq:zeta}
  G_0 \equiv G(I_0) = \zeta(\beta) ,
\end{equation}
where $\zeta = (\zeta^+ + \zeta^-)_{x=0}$, 
$${\zeta^\pm} = \mu^\pm  \frac{( 1 - r^\pm )}{(1 + r^\pm)},
$$ 
and ``$+$'' and ``$-$'' stand for the characteristics on the right and left sides of the waveguide, respectively, i.e. $\nu(x) = \nu^+(|x|)$, for $x>0$, and $\nu(x) = \nu^-(|x|)$, for $x<0$. 

Relation~(\ref{eq:zeta}) allows us to identify different types of nonlinear localized states. We notice that such localized states exist in the so-called {\nem waveguiding} regime ($G_0 > 0$), when $\zeta(\beta) > 0$. Additionally, spatial localization can occur in the {\nem anti-waveguiding} regime ($G_0 < 0$), provided $\zeta(\beta) < 0$. 

To study {\nem linear stability} of the localized solutions, we consider the evolution of small-amplitude perturbations of the localized state presenting the solution in the form
\begin{equation} \label{eq:lperturb}
 \psi (x,t) 
 = \left\{ u(x) + v(x) e^{i \Gamma z} 
                + w^{\ast}(x) e^{-i \Gamma^{\ast} z} 
   \right\} e^{i \beta z} ,
\end{equation}
and obtain the linear eigenvalue problem for small $v(x)$ and $w(x)$,
\begin{equation} \label{eq:Leigen}
  \begin{array}{l}
   { \displaystyle
      -(\beta+\Gamma) v + \frac{d^2 v}{d x^2} + \nu(x) v
   } \\*[9pt] { \displaystyle \qquad\qquad\qquad\qquad
      + \delta(x) \left[ G_1 v + (G_1 - G_0) w \right] = 0 ,
   } \\*[9pt] { \displaystyle
      -(\beta-\Gamma) w + \frac{d^2 w}{d x^2} + \nu(x) w
   } \\*[9pt] { \displaystyle \qquad\qquad\qquad\qquad
      + \delta(x) \left[ G_1 w + (G_1 - G_0) v \right] = 0 ,
   } \end{array}
\end{equation}
where $G_1 \equiv G_0 + I_0 G^{\prime}(I_0)$. The intensity $I_0$ is calculated for an unperturbed solution, and the prime stands for the spatial derivative.

We find that the localized eigenmode solutions of Eq.~(\ref{eq:Leigen}) exist only for the particular eigenvalues satisfying the solvability condition, $Y( \gamma ) = 0$~\cite{evans}, where
\[
Y( \gamma  ) =
   \left[ G_1 - \zeta(\beta+\Gamma) \right]
   \left[ G_1 - \zeta(\beta-\Gamma) \right]
   - \left( G_1 - G_0 \right)^2 .
\]
In general, the eigenmode solutions fall into one of the following categories:
(i)~{\nem internal modes} with real eigenvalues describe periodic oscillations (``breathing'') of the localized state, (ii)~{\nem instability modes} correspond to purely imaginary eigenvalues, and (iii)~{\nem oscillatory instabilities} can occur when the eigenvalues are complex.

To demonstrate the basic stability results, we consider a localized waveguide possessing a cubic nonlinear response,
  $G(I) = \alpha + \gamma I$.
Under proper scaling, the absolute value of the nonlinear coefficient~$\gamma$ can be normalized to unity, so that $\gamma=+1$ corresponds to {\nem self-focusing} and $\gamma=-1$ to {\nem self-defocusing} nonlinearity. We assume that at small intensities the waveguide has a higher refractive index, i.e. $\alpha>0$. Therefore, below we consider two qualitatively different examples corresponding to different signs of the nonlinear coefficient $\gamma$.

\subsection{Self-focusing nonlinearity}
            \label{sect:layer_sf}

\begin{figure}
\setlength{\epsfxsize}{8cm}
\vspace*{0mm}
\centerline{\mbox{\epsfbox{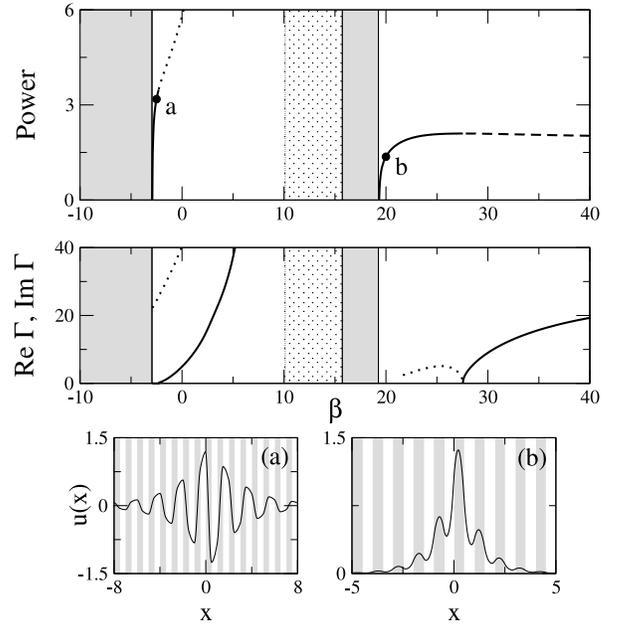}}}
\vspace*{0mm}
\caption{ \label{fig:pwr-sf-alp} \label{fig:bandgap}
Top: power vs. propagation constant for the nonlinear localized states: solid~--- stable, dashed~--- unstable, and dotted~--- oscillatory unstable.
Middle: real (dotted) and imaginary (solid) parts of the eigenvalues associated with the wave instability.
Shading marks ``waveguiding'' (white) and ``anti-waveguiding'' (dotted) localization regimes inside the band gaps.
Bottom:~the localized states corresponding to the marked points (a,b) in the top plot; shading marks the areas with smaller $\nu$.
The lattice parameters are $h=1$, $\nu(x)=0$ for $n-1/2 < x/h < n$, and $\nu(x)=30$ for $n < x/h < n+1/2$, where $n$ is integer, and $\alpha = 0.5$, $\gamma=1$.}
\end{figure}

We consider the properties of spatially localized waves supported by a thin-film waveguide with a self-focusing nonlinearity ($\gamma=+1$). Such waves do exist in the linear limit, and the corresponding propagation constant $\beta_b$ is defined by the equation $\zeta(\beta_b)=\alpha$. Such localized waves  correspond to the waveguiding regime (white regions in Fig.~\ref{fig:pwr-sf-alp}, top). 
We notice that the first band-gap (semi-infinite white region at $\beta > 19$) corresponds to the conditions of {\nem internal reflection} (IR) and, therefore, its diffraction properties should be similar to those of the conventional guided waves. Localized waves in this band resemble the guided waves modulated by a periodic structure [see Fig.~\ref{fig:bandgap}(b)]. Since an IR wave profile does not contain zeros, it is a fundamental eigenmode. Therefore, the conditions of the Vakhitov-Kolokolov (VK) stability theorem~\cite{VK} are satisfied, and the IR states are {\nem unstable} if and only if $d P / d \beta < 0$. At the critical point, $d P / d \beta = 0$, the linear eigenvalue passes through zero and becomes imaginary, as illustrated in Fig.~\ref{fig:pwr-sf-alp} (middle, $\beta \simeq 27.5$). 

\begin{figure}
\setlength{\epsfxsize}{8cm}
\vspace*{0mm}
\centerline{\mbox{\epsfbox{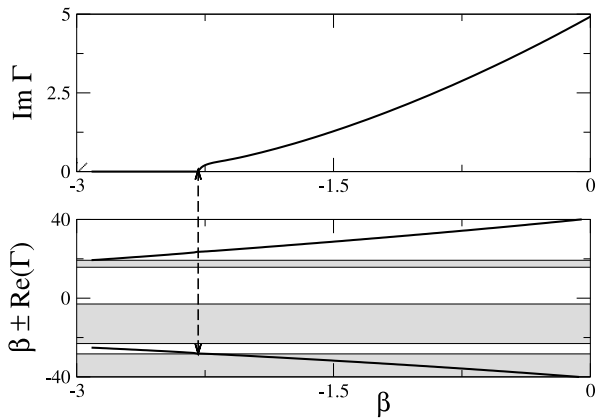}}}
\vspace*{-30mm}
\caption{ \label{fig:imode-sf-alp}
Example of a resonance that occurs between an internal mode of the localized state and a bang-gap edge, leading to an oscillatory instability.}
\end{figure}

In contrast, 
{\nem band-gaps} appear at smaller $\beta$ due to the resonant Bragg reflection (BR) by the periodic structure, so that the localized waves are similar to gap solitons composed of the mutually coupled backward and forward propagating waves [see Fig.~\ref{fig:bandgap}(a)]. For the BR states, the VK criterion provides only a necessary condition for stability, since the higher-order localized states can also exhibit oscillatory instabilities. Indeed, we notice that in the linear limit there always exists an internal mode corresponding to a resonant coupling between the BR and IR band-gaps, since $Y ( \beta_b^{\rm (IR)}-\beta_b^{\rm (BR)} ) \equiv 0$. We perform extensive numerical calculations and find that this mode leads to an oscillatory instability of BR waves when the value $(\beta-{\rm Re}\;\Gamma)$ moves outside the band gap; it occurs when the wave intensity exceeds a threshold value (see Figs.~\ref{fig:pwr-sf-alp} and~\ref{fig:imode-sf-alp}).

\subsection{Self-defocusing nonlinearity}
            \label{sect:layer_df}

\begin{figure}
\setlength{\epsfxsize}{8cm}
\vspace*{0mm}
\centerline{\mbox{\epsfbox{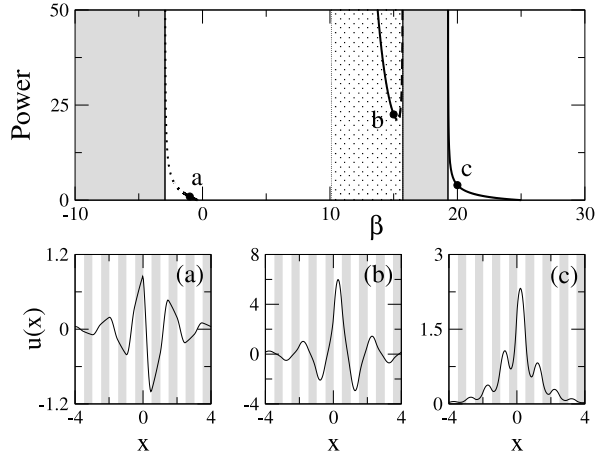}}}
\vspace*{-25mm}
\caption{ \label{fig:pwr-df-alp}
Top: power vs. propagation constant, and (a-c) the localized mode profiles for $\alpha = 5$, $\gamma = -1$. Notations are the same as in Fig.~\ref{fig:pwr-sf-alp}.}
\end{figure}

Additionally, we study the case of a thin-film linear waveguide ($\alpha>0$) possessing a self-defocusing nonlinearity ($\gamma=-1$). At small intensities, the waveguide supports linear guided waves in the waveguiding regime, in both IR and BR gaps, as shown in Fig.~\ref{fig:pwr-df-alp} (top, white regions). Performing the stability analysis, we find that IR waves are always {\nem stable}, while oscillatory instabilities appear for higher-order band-gap states. Since at higher intensities, i.e. for $I_0 > |\alpha/\gamma|$, the waveguide response changes its sign, a new type of localized waves can exist in the {\nem anti-waveguiding} regime (i.e. for the negative effective diffraction), bifurcating from the band-gap edge, as shown in Fig.~\ref{fig:pwr-df-alp} (top, dotted region). It is possible to demonstrate that for such waves {\nem the VK stability criterion becomes inverted}, i.e. the localized waves are unstable if  $d P / d \beta > 0$. This happens because the signs of both the nonlinear response and effective diffraction are altered compared to the IR waves supported by a self-focusing waveguide. All higher-order band-gap states exhibit oscillatory instabilities.

\section{An array of nonlinear waveguides}
         \label{sect:array}
\subsection{Model and Discrete Equations}
            \label{sect:array_model}

Now we analyze both localized and extended nonlinear waves, supported by a periodic array of thin-film waveguides with the Kerr-type nonlinear response. The corresponding model is a simplified version of the nonlinear layered medium usually studied in experiment~\cite{exp}. In our case, we assume again that the nonlinear layers are thin, so that the problem can be studied analytically, since the nonlinearity enters the periodic matching conditions only. 

To simplify our analysis further, we assume that the linear periodicity is associated only with the presence of the thin-film waveguides, and define the response function in the model Eq.~(\ref{eq:nls}) as follows
\begin{equation} \label{eq:nls_array_resp}
   {\cal F}(I; x) = \sum_n (\alpha + \gamma I) \delta(x - h n),
\end{equation}
where $h$ is the distance between the neighboring waveguides, and $n$ is integer. As above, the response of thin-film waveguides is approximated by the delta-functions, and the real parameters $\alpha$ and $\gamma$ describe both {\em linear} and {\em nonlinear} properties, respectively.  The nonlinear coefficient~$\gamma$ can be normalized to unity, so that $\gamma=+1$ corresponds to {\em self-focusing} and $\gamma=-1$ to {\em self-defocusing} nonlinearity. The linear coefficient ($\alpha>0$) defines the linear response (i.e. the response at the lower intensities), and  it characterizes the corresponding coupling strength between the waveguides.
 
As has been demonstrated in Ref.~\cite{our_pre},  the stationary wave profiles defined by Eqs.~(\ref{eq:lmode}) and (\ref{eq:u0_inh}) can be expressed in terms of the wave amplitudes at the nonlinear waveguides, $u_n = u(h n)$, which satisfy a stationary form of the discrete NLS equation:
\begin{equation} \label{eq:dnls}
  \eta U_n + ( U_{n-1} + U_{n+1} ) + \chi |U_n|^2 U_n = 0.
\end{equation}
Here $U_n = \sqrt{|\xi \gamma|} u_n$ are the normalized amplitudes, $\chi = {\rm sign}(\xi \gamma)$, and 
\begin{equation} \label{eq:dnls_param}
 \begin{array}{l} 
 { \displaystyle
   \eta = - 2\; {\rm cosh}( h \mu ) 
           + \alpha \xi,
 } \\*[9pt] { \displaystyle
   \xi = {\rm sinh}( h \mu ) / \mu, \;\;\; 
   \mu = \sqrt{\beta} .
 } \end{array}
\end{equation}
\begin{figure}[H]
\setlength{\epsfxsize}{7.5cm}
\mbox{\epsffile{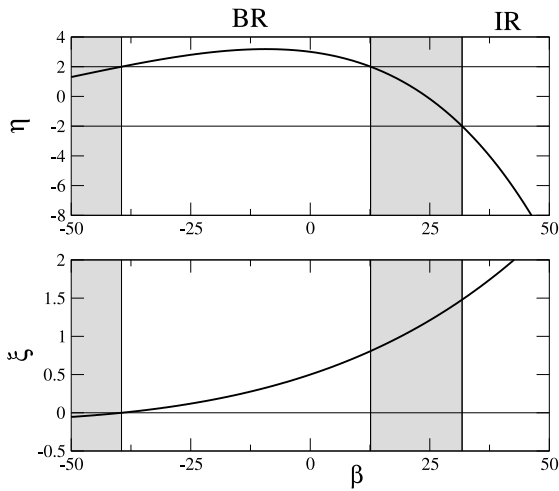}}
\vspace*{-15mm}
\caption{ \label{fig:dispers}
Characteristic dependencies of the parameters $\eta$ and $\xi$ on the propagation constant $\beta$. White shading marks band gaps.
The lattice parameters are $h=0.5$ and $\alpha=10$.
}
\end{figure}

Localized solutions of Eq.~(\ref{eq:dnls}) with exponentially decaying asymptotics can exist for $|\eta| > 2$, and this condition defines the spectrum {\em band-gap structure}. Characteristic dependencies of $\eta$ and $\xi$ vs. the propagation constant $\beta$ are presented in Fig.~\ref{fig:dispers}, where the band gaps are shown as blank stripes.  The first (semi-infinite) band gap corresponds to the total {\em internal reflection} (IR).
On the other hand, at smaller $\beta$ the {\em spectrum band gaps} appear due to the resonant Bragg-type reflection (BR) from the periodic structure. 
\subsection{Modulational instability}
            \label{sect:array_mi}

First, we analyze the properties of the simplest (periodic) solutions of the model~(\ref{eq:nls}) and~(\ref{eq:nls_array_resp}) with the equal intensities at the nonlinear layers, $I_0 = |u_n|^2 = {\rm const}$. Such Bloch-wave (BW) extended modes are characterized by the wave number in the first Brillouin zone, $-\pi \le K \le \pi$, so that $u_n = u_0 e^{i K n}$ and 
\begin{equation} \label{eq:BW_I0}
   I_0 = {-[2 \cos(K)+\eta]/(\gamma \xi)}. 
\end{equation}
These modes exist in the first transmission band which is shifted due to a nonlinear correction to the response function.

One of the main problems associated with the nonlinear BW modes is their {\em instability to periodic modulations} of a certain wavelength, known as {\em modulational instability}. In order to describe the stability properties of the periodic BW solutions, we analyse the linear evolution of weak perturbations in the form~(\ref{eq:lperturb}). 
Now, the functions $v(x)$ and $w(x)$ describe the small-amplitude perturbation corresponding to the eigenvalue $\Gamma$. Substituting Eq.~(\ref{eq:lperturb}) into the original model~(\ref{eq:nls}) and~(\ref{eq:nls_array_resp}), we obtain an eigenvalue problem which has the periodic solutions in the form of the Bloch functions, $v(x+h)=v(x) e^{i (q+K)}$ and $w(x+h) =  w(x) e^{i (q-K)}$, provided the following solvability condition is satisfied, 
\begin{equation} \label{eq:mi_cond}
 \begin{array}{l}
  {\displaystyle
    \left[ \eta(\beta+\Gamma) + 2 \gamma \xi(\beta+\Gamma) I_0 
           + 2 \cos(q + K) \right]
  } \\*[9pt] {\displaystyle
    \left[ \eta(\beta-\Gamma) + 2 \gamma \xi(\beta-\Gamma) I_0 
           + 2 \cos(q - K) \right]
  } \\*[9pt] {\displaystyle
    = \gamma^2 \xi(\beta+\Gamma) \xi(\beta-\Gamma) I_0^2.
  } \end{array}
\end{equation}
The spectrum of the possible eigenvalues $\Gamma$ is determined from the condition that the spatial modulation frequencies $q$, which are found from Eq.~(\ref{eq:mi_cond}), are all real.

\begin{figure}[H]
\setlength{\epsfxsize}{7.5cm}
\mbox{\epsffile{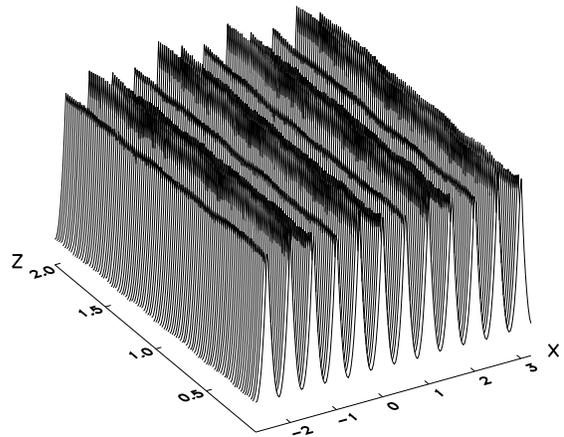}}
\vspace*{1mm}
\caption{ \label{fig:mi_bpm_unstag}
Development of modulational instability in a self-focusing medium for a slightly perturbed unstaggered BW solution with $I_0 \simeq 0.44$.  The lattice parameters are $\alpha=3$ and $h=0.5$, and $\gamma=+1$.
}
\end{figure}

In what follows, we consider two characteristic cases of the stationary BW profiles, when they are (i)~unstaggered ($K=0$) or (ii)~staggered ($K=\pi$). Modulational instability of the periodic BW solutions 
has been earlier studied for the Bose-Einstein condensates in optical lattices~\cite{mi_bec} in the mean-field approximation using the Gross-Pitaevskii equation, which is mathematically equivalent to Eq.~(\ref{eq:nls}) with ${\cal F}(I; x) = \nu(x) + \chi I$. It was demonstrated that the unstaggered solutions are always modulationally unstable in a self-focusing medium ($\chi=+1$), and they are stable in a self-defocusing medium ($\chi=-1$). It can be proven that the similar results are also valid for our model. We note that at small intensities the modulational instability in the self-focusing case corresponds to long-wave excitations, as illustrated in Fig.~\ref{fig:mi_bpm_unstag}.

\begin{figure}[H]
\setlength{\epsfxsize}{7.5cm}
\mbox{\epsffile{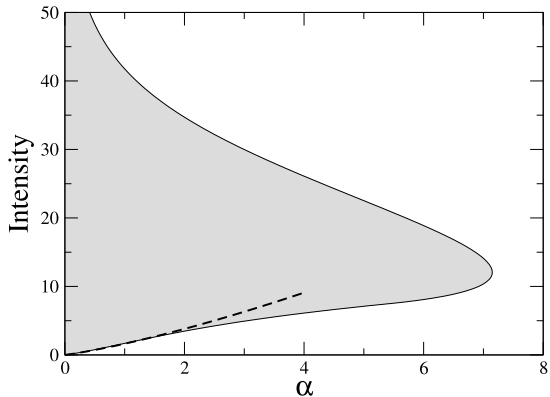}}
\vspace*{-25mm}
\caption{ \label{fig:mi_int}
Modulationally unstable staggered BW modes (gray shading) in a self-focusing medium, shown as the intensity ($I_0$) vs. the lattice parameter $\alpha$ (at $h=0.5$, $\gamma=+1$). Dashed line is the analytical approximation~(\ref{eq:mi_Icr}) for the low-intensity instability threshold.
}
\end{figure}

The staggered BW modes in a self-defocusing medium ($\chi=-1$) are always modulationally unstable~\cite{mi_bec}. On the other hand, the properties of such modes in a self-focusing medium ($\chi=+1$) are less trivial. We find that, in this case, the oscillatory instabilities (i.e. those with the complex $\Gamma$) can appear due to resonances between the modes of different bands. Such instabilities appear in a certain region of the wave intensities in the case of shallow modulations, when $\alpha$ is below a certain threshold value, as shown in Fig.~\ref{fig:mi_int}.  

It is interesting to compare these results with the results obtained in the framework of the continuum coupled-mode theory, valid for the case of a narrow band gap, i.e. for small $\alpha$ and small $I_0$. As was demonstrated earlier~\cite{mi_bragg}, in such a case the oscillatory instability appears above a certain critical  intensity, which is proportional to the band-gap width. In our case, we find the following asymptotic expression for the low-intensity instability threshold, 
\begin{equation} \label{eq:mi_Icr}
  \gamma I_0^{(cr)} \simeq \alpha 
                   + 2 \sqrt{2 h} \alpha^{3/2} / \pi 
                   + O(\alpha^{2}) .
\end{equation}
Since the band-gap width is $2 \alpha / h + O(\alpha^{3/2})$, we observe a perfect agreement with the results of the coupled-mode theory (see Fig.~\ref{fig:mi_int}, dashed line). 

In the limit of large $\alpha$ (or large $I_0$), the BW solution is composed of weakly interacting waves,  each localized at the individual waveguide. Then, the collective dynamics of the nonlinear modes in such a system can be studied with the help of the tight-binding approximation valid for the model of weakly coupled oscillators. This approach leads to the effective discrete NLS equation, which predicts the stability of the staggered modes in a self-focusing medium~\cite{mi_dnls}. Numerical and analytical results confirm that our solutions are indeed stable in the corresponding parameter regions.

\begin{figure}[H]
\setlength{\epsfxsize}{7.5cm}
\mbox{\epsffile{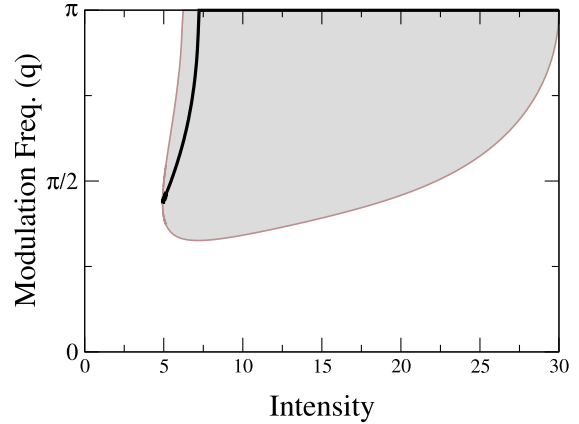}}
\vspace*{-25mm}
\caption{ \label{fig:mi_freq}
Unstable modulation frequencies (gray shading) vs. the intensity $I_0$ of the staggered BW modes ($\alpha=3$, $h=0.5$). Solid line shows the instability with the largest growth rate.
}
\end{figure}

\begin{figure}[H]
\setlength{\epsfxsize}{7.5cm}
\mbox{\epsffile{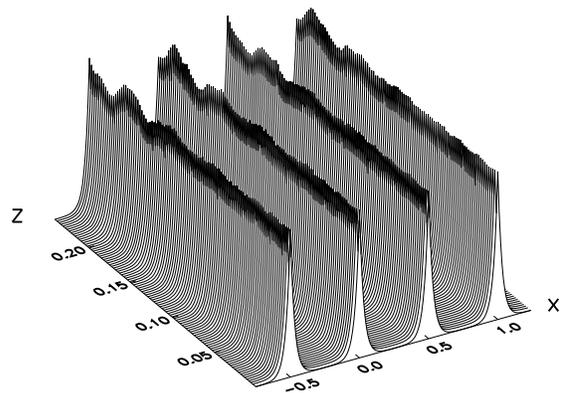}}
\vspace*{1mm}
\caption{ \label{fig:mi_bpm_stag}
Development of the instability-induced period-doubling modulations. Initial profile corresponds to a slightly perturbed staggered solution with $I_0 \simeq 29.87$. The parameters are the same as in Fig.~\ref{fig:mi_freq}.
}
\end{figure}

We find that in a self-focusing medium the staggered waves are always stable with respect to low-frequency modulations, see Fig.~\ref{fig:mi_freq}. However, at larger intensities unstable frequencies are shifted towards the edge of the Brillouin zone,  $q=\pi$, which has the largest value of the growth rate. The corresponding modulational instability manifests itself through the development of the period-doubling modulations, as shown in Fig.~\ref{fig:mi_bpm_stag}.

\subsection{Bright Spatial Solitons} \label{sect:bright}
\subsubsection{Odd and even localized modes} \label{sect:bright_slv}

Stationary localized modes in the form of discrete bright solitons can exist with the propagation constant inside the band gaps, when $|\eta|>2$. Additionally, it has been found that such solutions exist only for $\eta \chi < 0$. It follows from Eq.~(\ref{eq:dnls_param}), that $\xi > 0$ in the IR gap and the first BR gaps (see Fig.~\ref{fig:dispers}), so that the type of the nonlinear response is fixed by the medium characteristics, since $\chi = {\rm sign} \gamma$. Therefore, the self-focusing nonlinearity can support bright solitons in the IR gap (where $\eta<-2$), i.e. in the conventional {\em wave-guiding} regime. In the case of the self-defocusing response,  bright solitons can exist in the first BR gap, owing to the fact that the sign of the effective diffraction is inverted ($\eta>2$). In the latter case, the mode localization occurs in the so-called
{\em anti-waveguiding regime}.

Let us now consider the properties of two basic types of the localized modes: {\em odd}, centered at a nonlinear thin-film waveguide, and {\em even}, centered between the neighboring waveguides, so that $U_{|n|} = \chi^s U_{-|n|-s}$, where $s=0,1$, respectively.  For discrete lattices, such solutions have already been studied in the literature~(see, e.g., Ref.~\cite{Campbell}), and it has been  found that the mode profile is ``unstaggered'' (i.e. $U_n>0$) if $\eta<-2$. On the other hand, Eq.~(\ref{eq:dnls}) possesses a symmetry transformation $U_n \rightarrow (-1)^n U_n$, $\eta \rightarrow -\eta$, and  $\chi \rightarrow -\chi$, which means that the solutions become ``staggered'' at $\eta>2$. 

\begin{figure}[H]
\setlength{\epsfxsize}{7.5cm}
\mbox{\epsffile{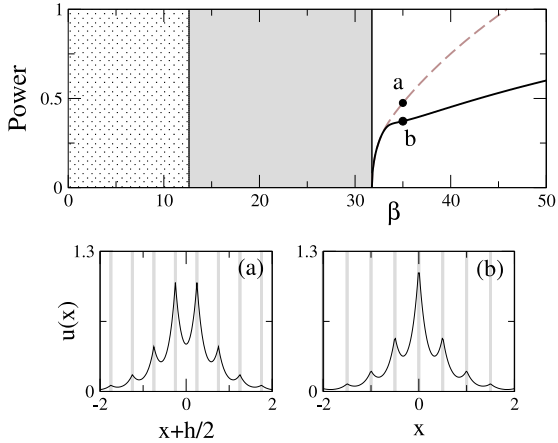}}
\vspace*{-20mm}
\caption{ \label{fig:pwr_sf}
Top: power vs. propagation constant for odd (black) and even (gray) localized modes in a self-focusing ($\gamma=+1$) regime: solid~--- stable, dashed~--- unstable, and dotted~--- oscillatory unstable.
Bottom:~the profiles of the localized modes corresponding to the marked points (a,b) in the top plot.
The lattice parameters are the same as in Fig.~\ref{fig:dispers}.
}
\end{figure}

To study {\em linear stability} of the localized modes, we consider the evolution of small-amplitude perturbations in the form~(\ref{eq:lperturb})
and,  using the original model of Eqs.~(\ref{eq:nls}) and~(\ref{eq:nls_array_resp}),  obtain the linear eigenmode problem for small $v(x)$ and $w(x)$:
\[
  \begin{array}{l}
   { \displaystyle
      -(\beta+\Gamma) v + \frac{d^2 v}{d x^2} 
   } \\*[9pt] { \displaystyle \qquad\qquad
      + \sum_n \left[(\alpha + 2 \gamma u_n^2) v + \gamma u_n^2 w \right]
                     \delta(x - h n) = 0 , 
   } \\*[9pt] { \displaystyle
      -(\beta-\Gamma) w + \frac{d^2 w}{d x^2} 
   } \\*[9pt] { \displaystyle \qquad\qquad
      + \sum_n \left[(\alpha + 2 \gamma u_n^2)  w + \gamma u_n^2 v \right]
                     \delta(x - h n) = 0 ,
   } \end{array}
\]
where $\Gamma$ is related to the instability growth rate (see Sec.~\ref{sect:layer_bands} above).

\begin{figure}[H]
\setlength{\epsfxsize}{7.5cm}
\mbox{\epsffile{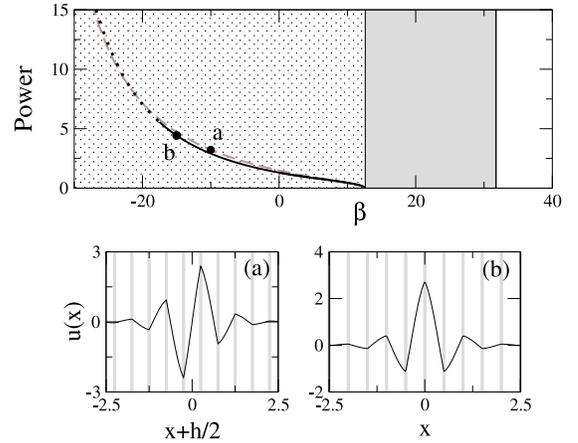}}
\vspace*{-20mm}
\caption{ \label{fig:pwr_df}
Top: power vs. propagation constant in the self-defocusing ($\gamma=-1$) regime.
Notations are the same as in Fig.~\ref{fig:pwr_sf}.
}
\end{figure}

Our analysis reveals that even modes are {\em always unstable} with respect to a translational shift along the $x$ axis. On the other hand, odd modes are always stable in the self-focusing regime [see Fig.~\ref{fig:pwr_sf}], but can exhibit {\em oscillatory instabilities} in the self-defocusing case when their power exceeds a certain critical value [see Fig.~\ref{fig:pwr_df}]. 

\subsubsection{Soliton bound states~--- ``twisted'' modes} \label{sect:bright_bs}

Due to a periodic modulation of the medium refractive index, solitons can form bound states~\cite{bound}. In particular, the so-called {\em ``twisted'' localized mode}~\cite{twisted} is a combination of two out-of-phase bright solitons~\cite{bound}. Such solutions do not have their continuous counterparts, and they can only exist when the discreteness effects are strong, i.e. for $|\eta| > \eta_{\rm cr}$. The properties of the twisted modes depend on the separation between the modes forming a bound state. We consider the cases of two lowest-order solution families (i)~of ``even'' type with zero nodes ($m=0$) in-between the peaks, and (ii)~of ``odd'' type with one node ($m=1$) in the middle, with $U_0 \equiv 0$. Then, the symmetry property is $U_{|n|+m} = - \chi^{m+1} U_{-|n|-1}$, and we find that $\eta_{\rm cr}(m=0) \simeq 3.32$ and $\eta_{\rm cr}(m=1) \simeq 2.95$.

\begin{figure}[H]
\setlength{\epsfxsize}{7.5cm}
\mbox{\epsffile{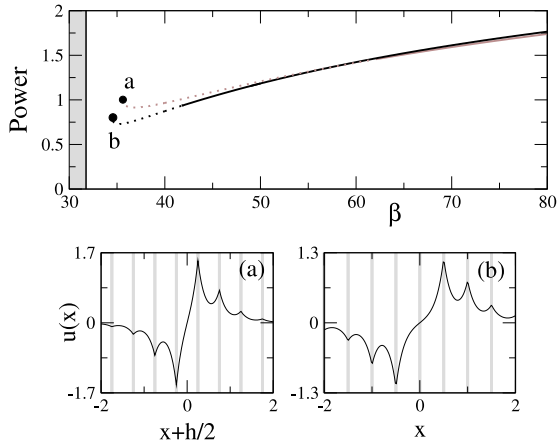}}
\vspace*{-20mm}
\caption{ \label{fig:pwr_sf_bright_tw}
Top: power vs. propagation constant for odd (black) and even (gray) twisted localized waves in a self-focusing ($\gamma=+1$) regime.
Notations are the same as in Fig.~\ref{fig:pwr_sf}.
}
\end{figure}

Stability properties of the twisted modes in the IR gap, under the self-focusing conditions, can be similar to those earlier identified in the framework of a discrete NLS model~\cite{twisted}. In the example shown in  Fig.~\ref{fig:pwr_sf_bright_tw}, the modes are stable at larger values of the propagation constant, and they become oscillatory unstable closer to the boundary of the existence region. Quite importantly, we see that the stability region is much wider in the case of the odd twisted modes due to a larger separation between the individual solitons of the bound state.

\begin{figure}[H]
\setlength{\epsfxsize}{7.5cm}
\mbox{\epsffile{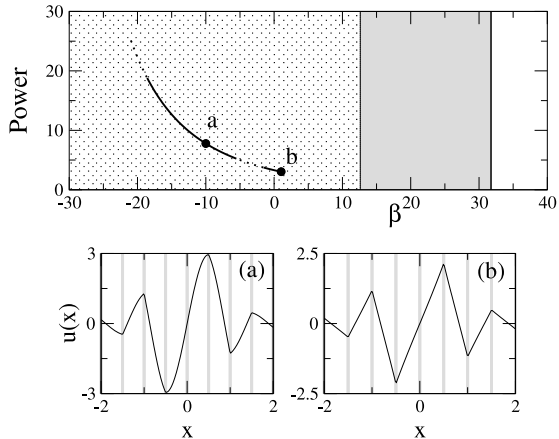}}
\vspace*{-20mm}
\caption{ \label{fig:pwr_df_bright_tw}
Top: power vs. propagation constant for the odd twisted localized waves in a self-defocusing ($\gamma=-1$) regime.
Notations are the same as in Fig.~\ref{fig:pwr_sf_bright_tw}.
}
\end{figure}

The characteristics of the twisted modes in the BR gap can differ substantially from the previous case. First, the value of $\eta$ is limited from above ($2 < \eta <\eta_{\rm max}$) and, therefore, some families of the twisted modes with $m < m_{\rm cr}$ may not exist. For example, for the medium parameters corresponding to Fig.~\ref{fig:dispers}(a), we have 
$\eta_{\rm cr}(m=1) < (\eta_{\rm max} \simeq 3.18 ) < \eta_{\rm cr}(m=0)$, 
so that $m_{\rm cr} = 1$. Under these conditions, the even modes with $m=0$ cannot exist in the BR regime. On the other hand, the odd modes with $m=1$ can exist, and they are stable in a wide parameter region, see Fig.~\ref{fig:pwr_df_bright_tw}.

\subsection{Dark Spatial Solitons} \label{sect:dark}

\begin{figure}[H]
\setlength{\epsfxsize}{7.5cm}
\mbox{\epsffile{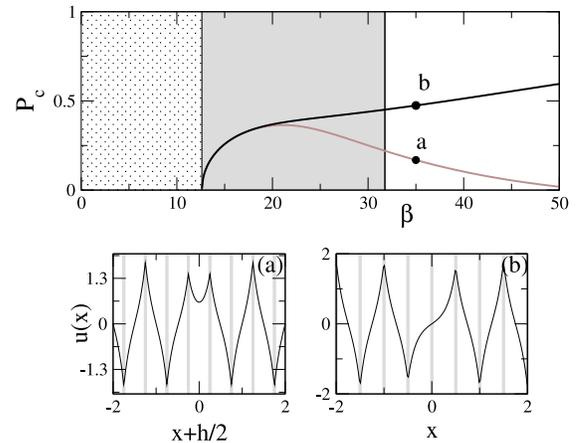}}
\vspace*{-20mm}
\caption{ \label{fig:pwr_sf_dark}
Top: complementary power vs. propagation constant for odd (black) and even (gray) dark localized solitons in a self-focusing ($\gamma=+1$) regime.
Notations are the same as in Fig.~\ref{fig:pwr_sf}.
}
\end{figure}

\begin{figure}[H]
\setlength{\epsfxsize}{7.5cm}
\mbox{\epsffile{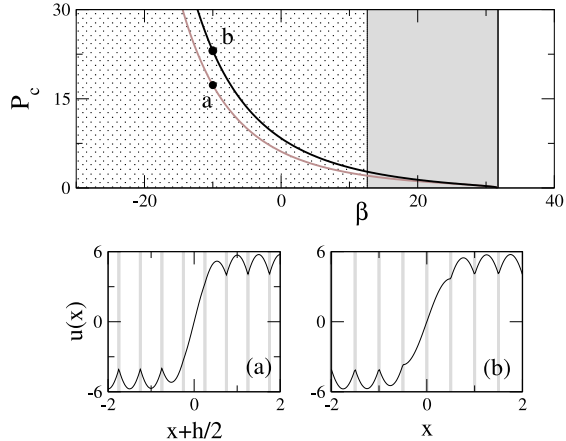}}
\vspace*{-20mm}
\caption{ \label{fig:pwr_df_dark}
Top: complementary power vs. propagation constant in the self-defocusing ($\gamma=-1$) regime. Notations are the same as in Fig.~\ref{fig:pwr_sf_dark}.
}
\end{figure}

Similar to the continuum NLS equation with self-defocusing nonlinearity~\cite{dark_review} or a discrete NLS model, our model supports localized modes on a Bloch-wave background, also called {\em dark solitons}. Such stationary localized modes appear at the band-gap edge where $\eta = 2 \chi$
(hereafter we consider only the case of a background corresponding to Bloch-wave solutions with $K=0,\pi$ introduced in Sec.~\ref{sect:array_mi}). As the background intensity increases, the band-gap edge shifts to match the dark soliton propagation constant $\beta$, and it follows from Eqs.~(\ref{eq:dnls_param}) and~(\ref{eq:BW_I0}) that $\eta(\beta; \tilde{\alpha}) = 2 \chi$, where $\tilde{\alpha} = (\alpha + \gamma u_{\infty}^2)$. Since these solutions correspond to the first transmission band, we have $\xi(\beta)>0$ (see Fig.~\ref{fig:dispers}), and it follows from Eq.~(\ref{eq:dnls_param}) that the propagation constant $\beta$ increases at higher intensities in a self-focusing medium ($\gamma>0$), and decreases in a self-defocusing medium
(see Figs.~\ref{fig:pwr_sf_dark} and~\ref{fig:pwr_df_dark}, top).  

Similar to the case of bright solitons discussed above, two basic types of dark spatial solitons can be identified as well, namely,  {\em odd localized modes} centered at a nonlinear thin-film waveguide, and {\em even localized modes} centered between the neighboring thin-film waveguides. Thus, all modes satisfy the symmetry condition, $U_{|n|+s} = - (-\chi)^{s+1} U_{-|n|-1}$, where $s=0,1$, respectively. The background wave is unstaggered if $\chi = -1$, and it is staggered for $\chi = +1$; the corresponding solutions  can be constructed with the help of a symmetry transformation $U_n \rightarrow (-1)^n U_n$. However, the stability properties of these two types of localized states can be quite different. Indeed, it has been demonstrated in Sec.~\ref{sect:array_mi} that the staggered background in a self-focusing medium ($\chi=+1$) can become unstable. On the contrary, the unstaggered background is always stable if $\chi=-1$.

Two types of the dark spatial solitons existing in our model are shown for both staggered and unstaggered background waves in Figs.~\ref{fig:pwr_sf_dark} and~\ref{fig:pwr_df_dark}, respectively. We characterize the family of dark solitons by the {\em complimentary power} defined as  
\[
  P_c = \lim_{n\rightarrow+\infty} \int_{-nh}^{+nh} 
         \left( |u(x+2 n h)|^2 - |u(x)|^2 \right) \; dx ,
\]
where $n$ is integer. 
The localized solutions shown in Fig.~\ref{fig:pwr_df_dark} are similar to those found earlier in Ref.~\cite{dark_NLS} in the context of the superflow structures in a periodic potential.

\section{Conclusion}

In the framework of a simplified model of a periodic layered medium, we 
have analyzed nonlinear guided waves and (bright, dark, and ``twisted") discrete spatial solitons of two types,  i.e. the nonlinear guided waves localized due to the total internal reflection and Bragg-like localized gap modes.  We have considered two specific geometries of the nonlinear periodic structures which may find their application in experimental realizations, namely, a single nonlinear layer embedded into a periodic linear medium, and a periodic array of identical nonlinear layers, the so-called one-dimensional nonlinear photonic crystal. We have analyzed the existence and stability of the nonlinear localized modes in both the models, and have described also modulational instability of homogeneous states in a periodic structure of the nonlinear layers.  In particular, we have discussed both similarity and difference with the models described by the discrete NLS equation, derived in the frequently used tight-binding approximation, and the results of the coupled-mode theory applicable for shallow modulations and a narrow gap.  We believe our results may be useful for other fields, such as the nonlinear dynamics of the Bose-Einstein condensates in optical lattices.

\section*{Acknowledgments}

The authors are indebted to Dr. O.~Bang, Prof. P.~L. Christiansen, and Prof. C.~M. Soukoulis for collaboration at the initial stage of this project, and to Prof. A.~A. Maradudin and Prof. Y.~Silberberg for useful and encouraging discussions. 
This work has been supported by the Performance and Planning Fund of the Institute of Advanced Studies, the Australian National University, and by the Australian Photonics Cooperative Research Center.

\end{multicols}
\end{document}